\documentclass[12pt,a4paper]{article}
\pdfoutput=1
\usepackage{graphicx}
\usepackage[T1]{fontenc}
\usepackage[sc,osf]{mathpazo}
\usepackage{a4wide}  
\usepackage{latexsym,amsthm,amsfonts,amsmath,mathrsfs,amssymb}
\usepackage[unicode,implicit]{hyperref}
\hypersetup{%
  pdftitle    = {Non-Perturbative decay of Non-Abelian hair}
  pdfkeywords = {Yang-Mills, black string, supergravity, supersymmetry,
    dimensional reduction, oxidation},
  pdfauthor   = {Pablo A. Cano, Tomas Ortin},
  plainpages  = true,
  colorlinks  = true,
  citecolor   = blue,
  urlcolor    = red,
  linkcolor   = black
}
\newcommand{\hepth}[1]{{\tt
\href{http://www.arXiv.org/abs/hep-th/#1}{hep-th/#1}}}
\newcommand{\grqc}[1]{{\tt
\href{http://www.arXiv.org/abs/gr-qc/#1}{gr-qc/#1}}}

\newcommand{\arxiv}[1]{{\tt arXiv:\href{http://www.arXiv.org/abs/#1}{#1}}}
\newcommand{\doi}[1]{{\tt DOI:\href{http://dx.doi.org/#1}{#1}}}
\makeatletter
\@addtoreset{equation}{section}
\makeatother

\begin{document}

\begin{flushright}
\small
IFT-UAM/CSIC-17-087\\
October 13\textsuperscript{th}, 2017\\
\normalsize
\end{flushright}

\vspace{1cm}

\begin{center}

{\Large {\bf {Non-perturbative decay of Non-Abelian hair}}}
 
\vspace{2.5cm}

\renewcommand{\thefootnote}{\alph{footnote}}
{\sl\large  Pablo A.~Cano}\footnote{E-mail: {\tt pablo.cano [at] uam.es}}
{\sl\large  and Tom\'{a}s Ort\'{\i}n}\footnote{E-mail: {\tt Tomas.Ortin [at] csic.es}}

\setcounter{footnote}{0}
\renewcommand{\thefootnote}{\arabic{footnote}}

\vspace{1cm}

{\it Instituto de F\'{\i}sica Te\'orica UAM/CSIC\\
C/ Nicol\'as Cabrera, 13--15,  C.U.~Cantoblanco, E-28049 Madrid, Spain}\\ 

\vspace{3cm}


{\bf Abstract}

\end{center}

\begin{quotation}
  {\small 
    We construct a solution of Heterotic supergravity which interpolates
    between two different AdS$_{3}\times S^{3}\times T^{4}$ geometries
    corresponding to the near-horizon limits of two 5-dimensional black holes,
    only one of which has non-Abelian hair. This solution can be used to
    estimate the amplitude of probability of the non-perturbative decay of the
    gauge 5-brane responsible for the non-Abelian hair into eight solitonic
    5-branes by evaluating its Euclidean action.  The Wick rotation of this
    solution poses several problems which we argue can be overcome by using a
    non-extremal off-shell (NEOS) deformation of the solution. This NEOS field
    configuration can be Wick rotated straight away and its Euclidean action
    can be computed for any value of the deformation parameter. The Euclidean
    result can then be anti-Wick-rotated and its extremal limit gives the
    Euclidean action of the original solution, which turns out to be one half
    of the difference between the entropies of the 5-dimensional black holes.
}
\end{quotation}

\newpage
\pagestyle{plain}


\section{Introduction}

During the last few years, a number of solutions of 4- and 5-dimensional
Super-Einstein-Yang-Mills theories\footnote{These are the minimal
  supersymmetrizations of the Einstein-Yang-Mills theories that admit
  supersymmetric black-hole solutions. Therefore, they are gauged
  supergravities with non-Abelian gauge groups and, due to the last condition,
  they must have at least 8 supercharges.}  describing extremal black holes,
strings and rings with different kinds of non-Abelian hair have been obtained in completely
analytic form
\cite{Hubscher:2008yz,Huebscher:2007hj,Meessen:2008kb,Bueno:2014mea,Meessen:2015nla,Meessen:2015enl,Ortin:2016bnl}. The
naive form of their masses and entropies is puzzling, since the non-Abelian
hair falls too fast at infinity to contribute to the mass but it is relevant
at the horizon and contributes to the entropy. Thus, there seemed to be an
infinite number of black holes with the same conserved charges at infinity but
completely different entropies. 

In order to clarify the situation, the embedding of a specially simple
5-dimensional black hole with non-Abelian hair in 10-dimensional Heterotic
Supergravity was recently studied in Ref.~\cite{Cano:2017qrq}. This embedding
leads to the identification of the physical parameters of the 5-dimensional
solution (Abelian charges and moduli) with the numbers of certain branes of
Heterotic String Theory, namely fundamental strings (F1s), solitonic 5-branes
(S5) and wave momentum in a compact direction (W). Furthermore, it was found
that a single gauge 5-brane \cite{Strominger:1990et} is responsible for all
the black hole's non-Abelian hair \cite{Cano:2017sqy} and that this gauge
5-brane contributes to the same 5-dimensional charge as the S5-branes with 8
units. Thus, this 5-dimensional charge, which occurs in the mass formula,
should be split into two different charges, Abelian and non-Abelian, both of
which contribute to the mass. In this way, all the branes of the solution contribute
to the mass, as expected, and the non-Abelian hair puzzle is solved by the
correct stringy identification of the charges. 

The solution to this puzzle poses new questions. Many of the non-Abelian
(single) black hole and string solutions we have constructed have very
interesting near-horizon geometries of a new kind that we have called
\textit{dumbbell} solutions in Ref.~\cite{Meessen:2017rwm} because they
interpolate between two Bertotti-Robinson-like spaces AdS$_{n}\times$ S$^{m}$
\cite{Bertotti:1959pf,Robinson:1959ev} with different radii. They were first
noticed in Ref.~\cite{Cano:2016rls}, but they arise for several values of $n$
and $m$ in near-horizon limits of non-Abelian black holes and black strings.
In particular, the near-horizon geometry of the simple non-Abelian 5-dimensional black hole studied in
Ref.~\cite{Cano:2017qrq} interpolates between two AdS$_{2}\times$ S$^{3}$
geometries with different radii which are found in two different limits,
$\rho\rightarrow 0,\infty$ of the radial coordinate. One of them contains the
contribution of the non-Abelian hair (\textit{i.e.}~the contribution of the
gauge 5-brane) while the other does not and it is just the AdS$_{2}\times$
S$^{3}$ one would obtain as the near-horizon geometry of a 3-charge Abelian
black hole.

The existence of this solution suggests its potential use to study the quantum
transition between one AdS$_{2}\times$ S$^{3}$ vacuum and the other by
Euclidean path integral methods,\footnote{See, for instance, the collection of
  reprints~\cite{Gibbons:1994cg}.} if a suitable instanton associated to this
Lorentzian solution can be found. As a matter of fact, it is easier to work
with and interpret the corresponding 10-dimensional solution of Heterotic
Supergravity that one obtains by oxidizing the 5-dimensional dumbbell
solution. In particular, as we are going to argue, in 10-dimensional language,
the transition between the two vacua can be interpreted as a transition
between a configuration that includes a gauge 5-brane and another in which
there is no gauge 5-brane but there are 8 additional S5-branes or, in other
words, the decay of a gauge 5-brane into 8 S5-branes (whose overall charge is
the same).

In spite of its simplicity (as compared to the 5-dimensional one), it is very
difficult to Wick-rotate the 10-dimensional solution to obtain the instanton
whose Euclidean action we need to evaluate. We are going to argue that the
most serious difficulties stem from the extremality of the solution. Indeed,
the direct evaluation of the Euclidean action of extremal black holes has
well-known problems \cite{Hawking:1994ii,Gibbons:1994ff} that do not arise
when one deals with the non-extremal solutions and finds their physical
quantities of interest (Hawking temperature and Bekenstein-Hawking entropy)
taking then the extremal limit of these quantities.

Here we propose to use a non-extremal deformation of the solution which is not
a solution: a non-extremal off-shell (NEOS) deformation since all one needs is
that the NEOS configurations interpolate between the same vacua as the
original solution (so they contribute to the path integral for the same
process), that they can be Wick-rotated and that they have finite Euclidean
action to take, afterwards, the extremal limit. There is no systematic
prescription to construct the NEOS configuration, but we manage to construct a
one-parameter family with just the right properties and we evaluate its
Euclidean action finding a result that we interpret physically as the
amplitude of probability of decay of a gauge 5-brane into 8 S5-branes in a
background containing a number of other branes. The value of the Euclidean action turns out to be one
half of the difference of the entropies of the non-Abelian and Abelian black
holes with those branes.

This paper is organized as follows: in Section~\ref{sec-solutions} we describe
the solutions we are going to work with. In Section~\ref{sec-tunneling} we
compute the above-mentioned amplitude, setting up the calculation in
Section~\ref{sec-setup}, rewriting it in Section~\ref{sec-dual} to make the
Wick rotation easier, introducing the NEOS configuration in
Section~\ref{sec-neos} and computing its Euclidean action in
Section~\ref{sec-computation}. In Section~\ref{sec-discussion} we discuss our
results.

\section{Black holes with non-Abelian hair}
\label{sec-solutions}

In this work we are going to study solutions of 10-dimensional Heterotic
Supergravity with just one set of SU$(2)$ gauge fields. Its action, in the
string frame, is given by

\begin{equation}
\label{heterotic}
{S}
=
\frac{g_{s}^{2}}{16\pi G_{N}^{(10)}}
\int d^{10}x\sqrt{|{g}|}e^{-2{\phi}}
\left[
{R}
-4(\partial{\phi})^{2}
+\tfrac{1}{2\cdot 3!}{H}^{2}
-\alpha'{F}^{A}{F}^{A}
\right]\, ,
\end{equation}

\noindent
where the 2- and 3-form field strengths ${F}^{A}$ and ${H}$ are
defined as

\begin{eqnarray}
{F}^{A}
& = & 
d{A}^{A}+\tfrac{1}{2}\epsilon^{ABC}{A}^{B}\wedge{A}^{C}\, ,
\\
& & \nonumber \\
\label{Hdef}
{H} 
& = & 
d{B}
+2\alpha' \left({F}^{A}\wedge {A}^{A}
-\tfrac{1}{3!} \epsilon^{ABC} 
{A}^{A}\wedge{A}^{B}\wedge{A}^{C} \right)\, ,
\end{eqnarray}

\noindent
$\alpha'=l_{s}^{2}$ where $l_{s}$ is the string length and the 10-dimensional
Newton constant $G_{N}^{(10)}$ is given  in terms of this string length and
the string coupling constant $g_{s}$ by

\begin{equation}
\label{eq:GN}
G_{N}^{(10)}=8\pi^{6}g_{s}^{2} l_{s}^{8}\, .
\end{equation}

The string coupling constant $g_{s}$ is equal to the vacuum expectation value of
the exponential of the dilaton $g_{s}=<\!\! e^{{\phi}}\!\!>$. In
asymptotically-flat solutions, this should also be the value of the dilaton at
infinity, and, therefore, in these solutions $g_{s}=e^{{\phi}_{\infty}}$.

This action is part of the low-energy effective field theory action of any of
the two Heterotic Superstrings at first order in $\alpha'$ since SU$(2)$ is
contained in both of their gauge groups.  From the supersymmetry point of
view, this action is \textit{complete}, \textit{i.e.}~the bosonic part of a
complete locally supersymmetric action. There is, however, another term which
enters the action at first order in $\alpha'$, proportional to
${R}_{-}^{2}$ where ${R}_{-}$ is the Lorentz curvature 2-form of one
of the torsionful spin connections ${\Omega}_{\pm\,
  {\mu}}^{{a}{b}} ={\omega}_{{\mu}}^{{a}{b}}\pm
\tfrac{1}{2}{H}_{{\mu}}^{{a}{b}}$. Also the Bianchi identity
of the 3-form field strength ${H}$ has another term to first order in
$\alpha'$, proportional to $\mathrm{Tr}\left({R}_{-}\wedge
  {R}_{-}\right)$.  Introducing these terms alone would break the
supersymmetric completeness of the action (a quartic term would be required to
restore it \cite{Bergshoeff:1989de}) and this is the reason why we are not
including them because we rely on supersymmetric solution-generating
techniques to obtain solutions. However, in order to consider the solutions of
this action as legitimate solutions of the full Heterotic String effective
action to first order in $\alpha'$ expansion, one has to show that the
terms quadratic in the curvature, evaluated on the solutions, are much smaller
than those we have kept. In the solutions that we are going to consider
(Eq.~(\ref{eq:nhsolution})) the $\mathrm{Tr}\left({R}_{-}\wedge
  {R}_{-}\right)$ and other ${R}_{-}^{2}$ terms are of higher order in
$\alpha'$.\footnote{From the point of view of $\alpha'$ corrections, the vector fields 
in the Heterotic action may be used to suppress the terms coming from the 
torsionful spin connection. In our case, the torsionful spin connection associated
 to the S5-brane equals the $SU(2)$ connection of the BPST instanton, so it is 
 natural to use non-Abelian fields. However, also Abelian bundles at large charge
  have been used in the literature in order to suppress the $R_{-}^2$ terms
   \cite{Halmagyi:2016pqu,Halmagyi:2017lqm}.}

In Ref.~\cite{Cano:2017qrq} we obtained the following solution of 
SU$(2)$ Heterotic Supergravity:

\begin{equation}
\label{eq:bhsolution}
\begin{aligned}
d{s}^{2}
& = 
\frac{2}{Z_{-}} du\left(dv-\frac{Z_{+}}{2}du\right)-\tilde
Z_{0}\left[d\rho^{2}+\rho^{2}d\Omega_{(3)}^{2}\right]-dy^{i}dy^{i}\, ,
\\
e^{2{\phi}}
& = 
e^{2{\phi}_{\infty}}\frac{\tilde Z_{0}}{Z_{-}}\, ,
\\
{B}
& = 
-\frac{1}{Z_{-}}dv\wedge du
-\frac{\tilde{Q}_{0}}{4} \cos{\theta}\, d\psi\wedge d\varphi\, ,
\\
{A}^{A}
& = 
-\frac{\rho^{2}}{\kappa^{2}+\rho^{2}}v^{A}_{L}\, ,
\end{aligned}
\end{equation}

\noindent
where $d\Omega_{(3)}^{2}$ is the metric of the unit, round $S^{3}$,
$v^{A}_{L}$ are the 3 left-invariant Maurer-Cartan 1-forms of SU$(2)$, the
coordinates $y^{i}$, $i=6,7,8,9$ parametrize a $T^{4}$ and the $Z$ functions
are given by

\begin{equation}
\begin{aligned}
\tilde{Z}_{0}
=
1+\frac{\tilde{Q}_{0}}{\rho^{2}}
+8\alpha'\frac{\rho^{2}+2\kappa^{2}}{(\kappa^{2}+\rho^{2})^{2}}\,
,
\hspace{2cm}
Z_{\pm}
=
1+\frac{Q_{\pm}}{\rho^{2}}\, ,
\end{aligned}
\end{equation}

\noindent
where, in their turn, $\kappa$ is the size parameter of a SU$(2)$ BPST
instanton and the charges $\tilde{Q}_{0},Q_{-}$ and $Q_{+}$ are related,
respectively, to the number of solitonic five-branes, $N_{S5}$, the number of
fundamental strings, $N_{F1}$, and the number of units of momentum flowing
along the compact direction $u$ of radius $R_{z}$, $N_{W}$, by

\begin{equation}
\label{eq:QversusN}
\tilde{Q}_{0}=l_{s}^{2} N_{S5}\, ,
\hspace{1cm}
Q_{-}=l_{s}^{2}g_{s}^{2} N_{F1}\, ,
\hspace{1cm}
Q_{+}=\frac{l_{s}^{4}g_{s}^{2}}{R_{z}^{2}} N_{W}\, .
\end{equation}

Apart from the $S5,F1$ and $W$ (``Abelian'') constituents, there is a single
\textit{gauge 5-brane} $N_{G5}=1$ sourced by the SU$(2)$ instanton
\cite{Strominger:1990et}.

When compactified on $T^{4}\times S^{1}$, this solution describes a
five-dimensional black hole with non-Abelian hair, whose entropy and mass read

\begin{eqnarray}
S 
& = & 
2\pi \sqrt{N_{S5}N_{F1}N_{W}}\, ,
\\
& & \nonumber \\
M 
& = & 
\frac{R_{z}^{2}}{l_{s}^{2}g_{s}^{2}}\left(N_{S5}+8N_{G5}\right)
+\frac{R_{z}}{l_{s}^{2}}N_{F1}+\frac{1}{R_{z}}N_{W}\, .
\end{eqnarray}

The objects we have referred to as ``Abelian'' source Abelian 1-forms in the
5-dimensional theory and their charges contribute both to the mass and
entropy. The non-Abelian gauge 5-brane manifests itself as a globally regular
gravitating instanton \cite{Cano:2017sqy} which contributes to the mass as 8
solitonic 5-branes would, but does not contribute to the entropy at all. This
makes this solution less thermodynamically favored than another one with
$N_{S5}'=N_{S5}+8$ solitonic 5-branes and no gauge 5-branes, which would have
exactly the same mass, the same Abelian charges and moduli at infinity but
larger entropy $S'=2\pi \sqrt{(N_{S5}+8)N_{F1}N_{W}}>S$, suggesting that the
spontaneous decay of a gauge 5-brane into 8 solitonic 5-branes is
thermodynamically possible.

However, the decay of a gauge 5-brane into 8 solitonic 5-branes can never take
place perturbatively, as the SU$(2)$ instanton is protected by topology and it
can only occur non-perturbatively, by quantum tunneling.

In order to study this decay, it is convenient to consider a related solution,
obtained by removing the $1$'s from the functions $Z_{0\pm}$, which can be
seen as the near-horizon limit of the above solution. This solution reads
explicitly

\begin{equation}
\label{eq:nhsolution}
\begin{aligned}
d{s}^{2}
& = 
\frac{2\rho^{2}}{Q_{-}} dudv-\frac{Q_{+}}{Q_{-}}du^{2}
-R^{2}\left(\frac{d\rho^{2}}{\rho^{2}}+d\Omega_{(3)}^{2}\right)-dy^{i}dy^{i}\,
,
\\
e^{2{\phi}}
& = 
e^{2{\phi}_{\infty}}\frac{R^{2}}{Q_{-}}\, ,
\\
{B} 
& = 
-\frac{\rho^{2}}{Q_{-}}dv\wedge du
-\frac{\tilde{Q}_{0}}{4} \cos{\theta} d\psi\wedge d\varphi\, ,
\\
{A}^{A} 
& = 
-\frac{\rho^{2}}{\kappa^{2}+\rho^{2}}v^{A}_{L}\, ,
\end{aligned}
\end{equation}

\noindent
where $R^{2}$ is the function

\begin{equation}
R^{2} = 
\tilde{Q}_{0}
+8\alpha'\frac{\rho^{2}(\rho^{2}+2\kappa^{2})}{(\kappa^{2}+\rho^{2})^{2}}\, .
\end{equation}

In the absence of non-Abelian fields, this solution would just be
AdS$_{3}\times$ S$^{3}\times$ T$^{4}$, globally. This is the near-horizon
geometry of the S5-F1-W brane configuration.  However, the above solution,
with the non-Abelian fields switched on, interpolates between two
AdS$_{3}\times$ S$^{3}\times$ T$^{4}$ geometries of different
radii:\footnote{The metric of the AdS$_{3}$ factor appears in a somewhat
  unconventional form
\begin{equation}
ds_{\rm AdS_{3}}^{2}
= 
\frac{2\rho^{2}}{Q_{-}} dudv-\frac{Q_{+}}{Q_{-}}du^{2}
-R_{0,\, \infty}^{2}\frac{d\rho^{2}}{\rho^{2}}\, ,
\end{equation}
but it can be checked that the Riemann curvature tensor corresponds to an
AdS$_{3}$ space of radius $R_{0,\infty}$.}

\begin{itemize}
\item In the $\rho\rightarrow 0$ limit the squared radius of the AdS$_{3}\times$
  S$^{3}$ factor\footnote{The radius of the AdS$_{3}$ and S$^{3}$ factors are
    equal, and we refer to this common value as the radius of the product
    geometry.}  is $R_{0}^{2}=\tilde{Q}_{0}$.
\item In the $\rho\rightarrow \infty$ limit the squared radius is
  $R_{\infty}^{2}=\tilde{Q}_{0}+8\alpha'$.
\end{itemize}

Furthermore, the gauge fields are also different in these two limits: 

\begin{itemize}
\item In the $\rho\rightarrow 0$ limit ${A}^{A}_{0}=0$.
\item In the $\rho\rightarrow \infty$ limit ${A}^{A}_{\infty}=-v^{A}_{L}$,
  which is a pure gauge configuration. 
\end{itemize}

In order to compare the two limits, we must gauge-transform
${A}^{A}_{\infty}$ so that it also vanishes
identically, ${A}^{A\, \prime}_{\infty}=0$. After this gauge
transformation, the 2-form ${B}$, which transforms simultaneously via
Nicolai-Townsend transformations due to the presence of the Chern-Simons
3-form, takes the form

\begin{equation}
{B}'
=
-\frac{\rho^{2}}{Q_{-}}dv\wedge du
-\frac{\tilde{Q}_{0}+8\alpha'}{4} \cos{\theta} d\psi\wedge d\varphi\, ,
\end{equation}

\noindent
which, on account of the first of Eqs.~(\ref{eq:QversusN}), tells us that the
asymptotic geometry contains $N_{S5}'=N_{S5}+8$ S5-branes. 

We conclude that the complete solution Eq.~(\ref{eq:nhsolution}) can be
interpreted as an interpolation between the near-horizon geometries of a
configuration with $N_{S5}$ S5-branes and $N_{G5}=1$ gauge 5-brane and another
configuration with $N_{S5}'=N_{S5}+8$ S5-branes and $N_{G5}=0$ gauge 5-branes.

\section{Tunneling amplitude from the Euclidean path integral}
\label{sec-tunneling}

\subsection{Setting up the calculation}
\label{sec-setup}

According to the Euclidean path integral approach, given an initial and a
final state at fixed Euclidean times, the transition probability amplitude between
them is given by

\begin{equation}
\label{eq:pathintegral}
\mathcal{Z}=\int \mathcal{D}[\Psi]e^{-S_{E}[\Psi]}\, ,
\end{equation}

\noindent
where is the integral is taken over all the Euclidean field configurations
$\Psi$ which satisfy the boundary conditions associated to the given initial
and final states and $S_{E}[\Psi]$ is their Euclidean action. This probability
can be well approximated by

\begin{equation}
\mathcal{Z}\sim e^{-S_{E}[\Psi_{0}]}\, ,
\end{equation}

\noindent
for a classical solution $\Psi_{0}$ with the given boundary conditions and
finite Euclidean action, \textit{i.e.}~an instanton.  In some cases (when the
initial and final states are vacua) this probability can be interpreted as
the decay rate of a metastable vacuum into a more stable one.

The simplest prescription to obtain a Euclidean solution is to Wick-rotate
($t=-i\tau$) a Lorentzian one. However, when applied to non-trivial field
configurations (non-static metrics, for example) this naive prescription fails
to give real solutions of Euclidean signature unless some parameters of the
solution are analytically continued into the complex domain (see, for
instance, the seminal Ref.~\cite{Gibbons:1976ue}). These Euclidean solutions
can be thought of as real sections of a complexified solution obtained by
analytical continuation, but their existence is by no means guaranteed and, in
general, finding a real solution of Euclidean signature associated to a
Lorentzian one is a well-known and complicated problem.\footnote{See
  \textit{e.g.}~Ref.~\cite{Visser:2017atf} and references therein.}

Here, we would like to find a real Euclidean solution associated to the
Lorentzian dumbbell solution described in Eqs.~(\ref{eq:nhsolution}) of the
previous section. Such a solution, if of finite Euclidean action, could be
interpreted as an instanton interpolating between the two vacua
$N_{S5},N_{G5}=1,N_{F1},N_{W}$ and $N_{S5}+8,N_{G5}=0,N_{F1},N_{W}$ and the
(minus) exponential of its Euclidean action would give the probability of
decay from one vacuum to the other. Predictably, after the preceeding
discussion, in the search for this real Euclidean solution we are going to
meet several problems that we are going to try to solve.

The first problem arises in the Wick rotation of the Kalb-Ramond 2-form
${B}$, which makes the electric part purely imaginary. This is usually
dealt with by Wick-rotating the ``electric charge,'' ($Q_{-}$, here) as well,
but this would prove fatal in this case because it would make the dilaton and
several components of the metric imaginary.

As we are going to explain, the root of this problem may lie in the
extremality of our solution.  Let us compare the Wick rotation of the solution
at hands with that of a more familiar solution: an electrically-charged
Reissner-Norstr\"om black hole.

In the non-extremal regime there is no problem with the simultaneous Wick
rotation of the time $t=-i\tau$ and the electric charge $q=-iq_{E}$ because in
the standard coordinates in which the metric takes the form

\begin{equation}
ds^{2} 
= 
\frac{(r-r_{+})(r-r_{-})}{r^{2}}  dt^{2} 
-\frac{r^{2}}{(r-r_{+})(r-r_{-})} dr^{2}
-r^{2}d\Omega^{2}_{(2)}\, ,
\end{equation}

\noindent
with $r_{\pm} = M\pm \sqrt{M^{2}-Q^{2}}$, the electric charge $Q$ only occurs
quadratically in the metric.  There is no extremal limit of this Euclidean
solution, though, as the Lorentzian extremality condition $M^{2}-Q^{2}=0$
becomes $M^{2}+Q^{2}_{E}=0$ and the near-horizon limit is always
$\mathbb{E}^{2}\times S^{2}$ (corresponding to the Lorentzian Rindler$\times
S^{2}$) and not $\mathbb{H}^{2}\times S^{2}$ (which would correspond to
AdS$_{2}\times$S$^{2}$). Of course, one can always take the extremal limit of
the physical quantities computed in the non-extremal case after they are
re-expressed in terms of the Lorentzian charges. This is how, typically, the
entropy and (vanishing) temperature of extremal black holes are computed in
the Euclidean approach because the direct Wick rotation and the computation of
the Euclidean action of the extremal solution present very serious
problems.\footnote{The Wick rotation of the extremal Kerr black hole provides
  another, slightly different, example of the same problem which can only be
  solved by working with non-extremal Kerr black holes, which can be
  Wick-rotated consistently if one also rotates the angular momentum, and then
  taking the extremal limit of the results expressed in terms of the Lorentzian
  variables. }

Let us start with the problems presented by the direct Wick rotation of the
extremal solution.

First of all, if one tries to Wick-rotate directly the extremal Lorentzian
solution in which $r_{+}=r_{-}=M=\pm Q$ one finds that one has to Wick-rotate
the mass as well, losing the reality of the metric. In our case there seems to
be no way to make all the Wick-rotated fields real (specially the metric, due
to its complicated form) simultaneously no matter how we treat the parameters
of the solution.

There is a way out in the context of the 4-dimensional Maxwell-Einstein
theory: one can dualize the electric charge into a magnetic charge, which
needs not be Wick-rotated. Dyonic solutions such as Eqs.~(\ref{eq:nhsolution})
can be more difficult to rotate into a purely magnetic solution but we can
split the 2-form into its electric and a magnetic parts and dualize only the
electric one obtaining two magnetic fields (a 2-form and a 6-form) which do
not need to be Wick-rotated. We will explain how to do this in detail later
but we can advance that this trick turns out to be only good enough to keep
the solution real for $Q_{+}=0$. This strongly suggests that we should try to
work with a non-extremal Euclidean solution and then take the extremal limit
of the Lorentzian results.

On top of the problems related to the Wick rotation there is another problem
that seems to affect extremal solutions only and which supports the need of
working with non-extremal solutions. As shown in
Refs.~\cite{Hawking:1994ii,Gibbons:1994ff}, a direct calculation of the
entropy of the extremal Reissner-Nordstr\"om black hole within the Euclidean
approach gives a result ($S_{\rm BH}=0$) which differs from the extremal limit
of the entropy of the non-extremal black holes, which is the same as the value
obtained by counting microstates in the String Theory context
\cite{Strominger:1996sh,Callan:1996dv,Maldacena:1996ky}. The technical reason
is the existence of an inner boundary in the extremal Euclidean solution which
does not exist in the non-extremal family of solutions for any value of the
physical parameters.

The need to work with a non-extremal solution raises another problem, because
the non-extremal version of the black-hole solution Eqs.~(\ref{eq:bhsolution})
is not known and it has been argued that it may not exist. We are going to
circumvent this problem by constructing a 1-parameter non-extremal deformation
of the solution Eqs.~(\ref{eq:nhsolution}) which \textit{is not a solution of
  the equations of motion but interpolates between the same two vacua as the
  extremal solution}. This \textit{non-extremal off-shell (NEOS)} deformation
can be understood as a mere regularization procedure or as a computation of
the action over an off-shell family of field configurations that contribute to
the path integral in Eq.~(\ref{eq:pathintegral}) because they have the
boundary conditions demanded in this case. The extremal limit is, at the same
time an extremum of the action because it is a solution of the classical
equations of motion and, clearly, it makes sense to compute the action over
the complete family of configurations and then take the extremal limit.

In the rest of this section we are going to carry out the program explained
above. First of all, we are going to dualize the electric part of the
Kalb-Ramond 2-form into a magnetic 6-form. This has to be done in the action
and in the solution simultaneously. Next, we will make a first attempt at the
Wick rotation and we will see that for $Q_{+}\neq 0$ we need the NEOS
deformation. Finally, we will compute the Euclidean action for this family of
field configurations, taking into account all the boundary terms.

\subsection{Dual action and solution}
\label{sec-dual}

In order to dualize the electric part of the Kalb-Ramond 2-form ${B}$ we
replace it by the sum of a pair of 2-forms $ {B}_{1}+{B}_{2}$ such
that ${H}_{1}\cdot {H}_{2}=0$ and then dualize the second into a
6-form $\tilde{{B}}_{2}$ with 7-form field strength $\tilde{{H}}_{2}=
\star e^{-2{\phi}} {H}_{2}$ such that ${H}_{1}\wedge
\tilde{{H}}_{2}=0$. The resulting action is\footnote{In the process of
  dualization a boundary term is also generated, which is not shown here
  because it does not change the equations of motion, but which will be taken
  into account in the computation of the Euclidean action.}

\begin{equation}
\label{heterotic2}
\begin{aligned}
{S} 
& = \frac{g_{s}^{2}}{16\pi G_{N}^{(10)}}\int d^{10}x\sqrt{| g|}
  \left\{ e^{-2{\phi}} \left[
      {R}-4(\partial{\phi})^{2}+\frac{1}{2\cdot 3!}{H}^{2}
      -\alpha'{F}^{A}{F}^{A} \right] 
\right.
\\
& \\
&
\left.
+\frac{e^{2{\phi}}}{2\cdot 7!}\tilde{H}^{2} 
-\frac{\alpha'}{2\cdot 6! \sqrt{| g|}} 
\epsilon^{\mu_{1}\cdots\mu_{6}\alpha\beta\gamma\delta}
\tilde{{B}}_{\mu_{1}\cdots\mu_{6}}
{F}^{A}{}_{\alpha\beta}{F}^{A}{}_{\gamma\delta}
\right\}\, , \\
\end{aligned}
\end{equation}

\noindent
and any solution of this action satisfying the constraint ${H}\wedge
\tilde{{H}}=0$ is a solution of the original Heterotic Supergravity with

\begin{equation}
{H}={H}+e^{2{\phi}}\star\tilde{{H}}\, .
\end{equation}

We can reexpress the solution Eqs.~(\ref{eq:nhsolution}) as the following
purely magnetic solution of the above action whose Wick rotation is
potentially simpler:

\begin{equation}
\label{Solution2}
\begin{aligned}
d{s}^{2}
& = 
\frac{2\rho^{2}}{Q_{-}} dudv-\frac{Q_{+}}{Q_{-}}du^{2}
-R^{2}\left(\frac{d\rho^{2}}{\rho^{2}}+d\Omega_{(3)}^{2}\right)
-dy^{i}dy^{i}\, ,
\\
e^{2{\phi}}
& = 
e^{2{\phi}_{\infty}}\frac{R^{2}}{Q_{-}}\, ,
\\
{B}
& = 
-\frac{\tilde{Q}_{0}}{4} \cos{\theta} d\psi\wedge d\varphi\, ,
\\
\tilde{{B}}
& = 
-\frac{e^{-2{\phi}_{\infty}}Q_{-}}{4} \cos{\theta} 
d\psi\wedge d\varphi\wedge dy^{6}\wedge dy^{7}\wedge dy^{8}\wedge dy^{9}\, ,
\\
{A}^{A}
& = 
-\frac{\rho^{2}}{\kappa^{2}+\rho^{2}}v^{A}_{L}\, .
\end{aligned}
\end{equation}

In order to proceed with Wick rotation, we first need to identify an
appropriate time coordinate. Let us first begin with the case in which there
is no momentum, $Q_{+}=0$. In this case, we can make the change of variables
$v=(t-x)/\sqrt{2}$, $u=(t+x)/\sqrt{2}$, which makes
$2dudv=dt^{2}-dx^{2}$. Then we substitute $t=-i\tau$, and redefine with a
global sign the metric to get a positive-definite Euclidean metric (the rest
of fields are unaffected)

\begin{equation}
d{s}_{E}^{2}
=
\frac{\rho^{2}}{Q_{-}} (d\tau^{2}+dx^{2})
+R^{2}\left(\frac{d\rho^{2}}{\rho^{2}}+d\Omega_{(3)}^{2}\right)+dy^{i}dy^{i}\, .
\end{equation}

This solution interpolates between two $\mathbb{H}^{3}\times$ S$^{3}$
geometries of radii $R_{0}$ and $R_{\infty}$ and is a gravitational instanton
which represents a tunneling history in which one $\mathbb{H}^{3}\times$
S$^{3}$ vacuum decays into another one of larger radius, or, according to the
previous discussion, a history in which a gauge 5-brane decays into 8
solitonic 5-branes. We may just compute the Euclidean action for this solution
which is real but, as  we will see, it vanishes.

\subsection{The NEOS deformation and its Euclideanization}
\label{sec-neos}

This is closely related to the fact that the solution with $Q_{+}=0$ does not
produce a black hole in five dimensions. Indeed, it is known that the on-shell
Euclidean action for black hole solutions is related to the entropy of the
black hole. For $Q_{+}=0$ the entropy of the would-be black hole vanishes so
it is reasonable that the Euclidean action does so. It is necessary to have
$Q_{+}\neq 0$ in order to get a non-vanishing action.\footnote{The $Q_{+}=0$
  solution becomes singular in $d=5$, while, for $Q_{+}\neq 0$ it is
  AdS$_{2}\times$S$^{3}$, the near-horizon limit of a regular, extremal black
  hole.}

When $Q_{+}\neq 0$ things are more involved: after the change of variables
$v=(t-x)/\sqrt{2}$, $u=(t+x)/\sqrt{2}$

\begin{equation}
\label{Lorentzian}
d{s}^{2}
= 
\frac{1}{Q_{-}}\left[\left(\rho^{2}-\frac{Q_{+}}{2}\right)dt^{2}
-Q_{+}dt dx-\left(\rho^{2}+\frac{Q_{+}}{2}\right)dx^{2}\right]
-R^{2}\left(\frac{d\rho^{2}}{\rho^{2}}+d\Omega_{(3)}^{2}\right)
-dy^{i}dy^{i}\, ,
\end{equation}

\noindent
there is a crossed term in the metric $Q_{+}dtdx$ that becomes imaginary after
the Wick rotation unless we rotate $Q_{+}$ as well. But $Q_{+}$ occurs in more
places in the metric, which would become complex.

As explained before, this problem can be solved by using a 1-parameter $a$
NEOS deformation of the metric\footnote{The matter fields in the solution
  Eqs.~(\ref{Solution2}) are unaffected by this deformation.}

\begin{equation}
\label{NonExt}
\begin{aligned}
d{s}^{2}_{\rm NEOS}
= & 
\frac{1}{Q_{-}}
\left[
\left((\rho+\rho_{0})^{2}-\frac{Q_{+}}{2}\right)dt^{2}
-a dt dx -\left((\rho+\rho_{0})^{2}+\frac{Q_{+}}{2}\right)dx^{2}
\right]
\\
&
\\
&
-R^{2}
\left[
\frac{(\rho+\rho_{0})^{2}d\rho^{2}}{(\rho+\rho_{0})^{4}-\rho_{0}^{4}}
+d\Omega_{(3)}^{2}
\right]
-dy^{i}dy^{i}\, ,
\\
\end{aligned}
\end{equation}

\noindent
where

\begin{equation}
\rho_{0}^{2}\equiv \tfrac{1}{2}\sqrt{Q_{+}^{2}-a^{2}}\, ,
\end{equation}

\noindent
and $R(\rho)$ has the same form as before.

Even though the NEOS configuration (\ref{NonExt}) is not a solution in
general, it shares with the original extremal solution several interesting
properties. If we consider the case of pure gauge vector fields, so that
$R(\rho)$ is constant, it is a solution with the geometry AdS$_{3}\times
S^{3}\times T^{4}$, albeit in different coordinates. With non-trivial gauge
fields the metric interpolates between two  AdS$_{3}\times S^{3}\times T^{4}$
geometries with radii $R_{0}$ and $R_{\infty}$, so the NEOS configurations
contribute to the path integral that describes the transition between these
two vacua, and the $a=Q_{+}$ extremizes the Euclidean action.

Furthermore, for arbitrary values of $a$, this metric can be analytically
continued to a Euclidean metric by making $t=-i\tau$, $a=i \aleph$. In general,
it is not a solution, but when the ``extremality condition'' $a^{2}=Q_{+}^{2}$
is satisfied in the Lorentzian configuration, we recover the solution
(\ref{Lorentzian}).

As discussed above, we are going to Wick-rotate the above NEOS configuration
first, then we are going to  compute its Euclidean action and at the end  we
are going to take the $a^{2}\rightarrow Q_{+}^{2}$ limit in the result.

First, let us massage a bit the metric by making the change of coordinates

\begin{equation}
\begin{aligned}
t
= & 
\frac{\sqrt{Q_{+}/2+\rho_{0}^{2}}}{\sqrt{2}\rho_{0}}t'
-\frac{\sqrt{Q_{+}/2-\rho_{0}^{2}}}{\sqrt{2}\rho_{0}}y\, ,
\\
&
\\
x 
= &
\frac{\sqrt{Q_{+}/2+\rho_{0}^{2}}}{\sqrt{2}\rho_{0}}y
-\frac{\sqrt{Q_{+}/2-\rho_{0}^{2}}}{\sqrt{2}\rho_{0}}t'\, , 
\end{aligned}
\end{equation}

\noindent
which is a Lorentz boost. This sets the metric in a diagonal form

\begin{equation}
\begin{aligned}
d{s}^{2} 
= & 
\frac{1}{Q_{-}}
\left\{
\left[(\rho+\rho_{0})^{2}-\rho_{0}^{2}\right]dt'^{2}
-\left[(\rho+\rho_{0})^{2}+\rho_{0}^{2}\right]dy^{2}
\right\}
\\
&
\\
&
-R^{2}
\left[
\frac{(\rho+\rho_{0})^{2}d\rho^{2}}{(\rho+\rho_{0})^{4}-\rho_{0}^{4}}+d\Omega_{(3)}^{2}
\right]-dy^{i}dy^{i}\, .
\\
\end{aligned}
\end{equation}

In the $\rho\rightarrow 0$ limit, the $t'-\rho$ sector of the metric is just a
$1+2$-dimensional Rindler spacetime and, therefore, it corresponds to a
non-extremal horizon whose Hawking temperature we will compute later.

To complete the definition of this field configuration we have to determine
the period of the coordinate $y$, which is assumed to be compact in the
original solution we started with. By comparing the area of the horizon of
this metric and of the original one we conclude that\footnote{Actually, we
  only need to impose that the period of $y$ tends to this quantity in the
  limit $a^{2}\rightarrow Q_{+}^{2}$.}

\begin{equation}
\label{eq:Ry}
2\pi R_{y}
\equiv
\int dy
=
\frac{\sqrt{Q_{+}/2}}{\rho_{0}}\int du
=
\frac{\sqrt{Q_{+}/2}}{\rho_{0}}2\pi R_{z}\, .
\end{equation}

Then, we perform Wick rotation: $t'=-i\tau$, $a=i\aleph$ and, after an overall
change of sign, we get the Euclidean metric

\begin{equation}
\begin{aligned}
d{s}^{2}_{\rm ENEOS}
= &
\frac{1}{Q_{-}}
\left\{
\left[(\rho+\rho_{0})^{2}-\rho_{0}^{2}\right]d\tau^{2}+
\left[(\rho+\rho_{0})^{2}+\rho_{0}^{2}\right]dy^{2}
\right\}
\\
&
\\
&
+R^{2}
\left[
\frac{(\rho+\rho_{0})^{2}d\rho^{2}}{(\rho+\rho_{0})^{4}-\rho_{0}^{4}}+d\Omega_{(3)}^{2}
\right]
+dy^{i}dy^{i}\, ,
\\
\end{aligned}
\end{equation}

\noindent
where now

\begin{equation}
\rho_{0}^{2}=\tfrac{1}{2}\left[Q_{+}^{2}+\aleph^{2}\right]^{1/2}\, .
\end{equation}
 
Note that the $R=$ constant configurations (with pure gauge vector fields) are solutions (just as their Lorentzian
partners) and have the geometry of $\mathbb{H}^{3}\times$ S$^{3}\times$
T$^{4}$ with radius $R$.

In the $\rho\rightarrow 0$ limit, redefining the
radial coordinate $\rho\equiv \rho_{0} r^{2}/R_{0}^{2}$, the metric takes the form

\begin{equation}
\begin{aligned}
d{s}^{2}_{E}
=
\frac{1}{Q_{-}}
\left[2r^{2}\frac{\rho_{0}^{2}}{R_{0}^{2}}d\tau^{2}+2\rho_{0}^{2}dy^{2}\right]
+dr^{2}+R_{0}^{2}d\Omega_{(3)}^{2}+dy^{i}dy^{i}\, ,
\end{aligned}
\end{equation}

\noindent
and, to avoid a conical singularity at $r=0$, the Euclidean time $\tau$ must be
periodic with period $\beta$ so that the Hawking temperature (undoing the Wick
rotation of the $a$ parameter) is

\begin{equation}
\label{eq:TH}
T_{H}=
\beta^{-1}
=
\frac{1}{2\pi}
\sqrt{\frac{(Q_{+}^{2}-a^{2})^{1/2}}{\tilde{Q}_{0}Q_{-}}}\, ,
\end{equation}

\noindent
and vanishes in the  $a\rightarrow Q_{+}$ limit.

\subsection{Computation of the Euclidean action}
\label{sec-computation}

The complete Euclidean action that we want to compute is given by

\begin{equation}
\label{EuclideanAction}
\begin{aligned}
S_{E}
= & 
\frac{g_{s}^{2}}{16\pi G_{N}^{(10)}}
\int_{\mathcal{M}} d^{10}x\sqrt{|{g}_{E}|}
\left\{e^{-2{\phi}}\left[{R}-4(\partial{\phi})^{2}+\frac{1}{2\cdot3!}
{H}^{2}+\alpha'{F}^{A}{F}^{A}\right]
\right.
\\
& \\
&
\left.
+\frac{e^{2{\phi}}}{2\cdot 7!}\tilde{{H}}^{2} 
+\alpha'\frac{\epsilon^{\mu_{1}\cdots\mu_{6}\alpha\beta\gamma\delta}}{2\cdot 6! \sqrt{|{g}_{E}|}} 
\tilde{{B}}_{\mu_{1}\cdots\mu_{6}}
{F}^{A}{}_{\alpha\beta}{F}^{A}{}_{\gamma\delta}
\right\}\, 
\\
&
\\
&
+\frac{g_{s}^{2}}{8\pi G_{N}^{(10)}}
\int_{\partial\mathcal{M}}d^{9}x\sqrt{|h_{E}|}
\left[
-\frac{e^{2{\phi}}}{2\cdot 6!}n^{\mu} 
(\tilde{{H}}\cdot \tilde{{B}})_{\mu} -e^{-2{\phi}}K
\right]+c\, ,
\end{aligned}
\end{equation}

\noindent
where $K$ is the trace of the extrinsic curvature of the boundary
$\partial\mathcal{M}$, $c$ is a normalization constant which is fixed by the
criterium that $S_{E}=0$ when evaluated on the vacuum (taken to be $\mathbb{H}^{3}\times$ S$^{3}\times$
T$^{4}$ with radius $R_{\infty}$) and $(
\tilde{{H}}\cdot \tilde{{B}})_{\mu}\equiv
\tilde{{H}}_{\mu\nu_{1}\cdots \nu_{6}}\tilde{{B}}^{\nu_{1}\cdots
  \nu_{6}}$. This surface term appears when we dualize the Kalb-Ramond 2-form
${B}$. For the normalization constant, we will use the usual prescription

\begin{equation}
c
=
\frac{g_{s}^{2}}{8\pi G_{N}^{(10)}}
\int_{\partial\mathcal{M}}d^{9}x\sqrt{|h_{E}|}e^{-2{\phi}}K_{0}\, ,
\end{equation}

\noindent
where $K_{0}$ is the extrinsic curvature of the boundary when it is embedded
in the vacuum.

In order to evaluate the integrand of the action it is convenient to use the
equations of motion, but here we have to be very careful because we are not
dealing with a solution, and not all of them are satisfied. In particular,
the equation of motion of the dilaton is not satisfied and we have to add a ``source'' term
$\delta(\rho;\aleph)$

\begin{equation}
\label{dilaton}
e^{-2{\phi}}
\left[
{R}-4(\partial{\phi})^{2}+\frac{1}{2\cdot 3!}{H}^{2}
+\alpha'{F}^{A}{F}^{A}\right]
-4\nabla_{\mu}(e^{-2{\phi}}\partial^{\mu}{\phi})
-\frac{e^{2{\phi}}}{2\cdot 7!}\tilde{{H}}^{2}
=
\delta(\rho;\aleph)\, ,
\end{equation}

\noindent
which can be simply computed by plugging the fields in the l.h.s. 

We know, however, that $\delta=0$ for global $\mathbb{H}^{3}\times$
S$^{3}\times$ T$^{4}$ and also that $\lim_{\rho\rightarrow\infty}
\delta(\rho;\aleph)=0$, because the configuration we are considering
asymptotically tends to a solution. Indeed, $\delta$ decays so fast that the
integral

\begin{equation}
\label{eq:Deltaa}
\Delta(\aleph)
\equiv
\frac{1}{Q_{-}}\int_{0}^{\infty}d\rho(\rho+\rho_{0})
R^{4}(\rho)\delta(\rho;\aleph)\, ,
\end{equation}

\noindent
converges.

Using the dilaton equation, the Euclidean action Eq.~(\ref{EuclideanAction})
takes a simpler form:

\begin{equation}
\label{EuclideanAction2}
\begin{aligned}
S_{E}
= & 
\frac{g_{s}^{2}}{16\pi G_{N}^{(10)}}
\int_{\mathcal{M}} d^{10}x\sqrt{|{g}_{E}|}
\left\{
\frac{e^{2{\phi}}}{7!}\tilde{{H}}^{2} 
+\alpha'\frac{\epsilon^{\mu_{1}\cdots\mu_{6}\alpha\beta\gamma\delta}}{2\cdot 6! \sqrt{|{g}_{E}|}} 
\tilde{{B}}_{\mu_{1}\cdots\mu_{6}}
{F}^{A}{}_{\alpha\beta}{F}^{A}{}_{\gamma\delta}
+\delta(\rho;\aleph)
\right\}\, 
\\
&
\\
&
+\frac{g_{s}^{2}}{8\pi G_{N}^{(10)}}
\int_{\partial\mathcal{M}}d^{9}x\sqrt{|h_{E}|}
\left\{
n_{\mu} 
\left[2e^{-2{\phi}}\partial^{\mu}{\phi}
-\frac{e^{2{\phi}}}{2\cdot 6!}
(\tilde{{H}}\cdot \tilde{{B}})^{\mu} 
\right]
-e^{-2{\phi}}(K-K_{0})
\right\}\, .
\end{aligned}
\end{equation}

Next, we massage the first term in the integrand

\begin{equation}
\frac{e^{2{\phi}}}{7!}\tilde{{H}}^{2} 
= 
\frac{e^{2{\phi}}}{6!}\tilde{{H}}^{\mu_{1}\cdots\mu_{7}}
\nabla_{\mu_{1}}\tilde{{B}}_{\mu_{2}\cdots\mu_{7}} 
= 
\frac{1}{6!}
\nabla_{\mu}\left[e^{2{\phi}}\left(\tilde{{H}}\cdot\tilde{{B}}\right)^{\mu}\right]
-\frac{1}{6!}\nabla_{\rho}\left(e^{2{\phi}}\tilde{{H}}^{\rho\mu_{1}\cdots\mu_{6}}\right)
\tilde{{B}}_{\mu_{1}\cdots\mu_{6}}\, .
\end{equation}

\noindent
The first term cancels identically the $\tilde{{H}}\cdot\tilde{{B}}$
boundary terms and the second combines with the
$\tilde{{B}}{F}{F}$ term into a term proportional to the equation
of motion of $\tilde{{B}}$,\footnote{This term has the form 
  \begin{equation}
  -\frac{1}{6!}\tilde{{B}}_{\mu_{1}\cdots\mu_{6}}
\left[
\nabla_{\rho}\left(e^{2{\phi}}\tilde{{H}}^{\rho\mu_{1}\cdots\mu_{6}}
\right)
-\alpha' 
\frac{\epsilon^{\mu_{1}\cdots\mu_{6}\alpha\beta\gamma\delta}}{2 
\sqrt{|{g}_{E}|}}
{F}^{A}{}_{\alpha\beta}{F}^{A}{}_{\gamma\delta}
\right]\, .  
  \end{equation}
}which happens to be identically satisfied,\footnote{Incidentally, this is the
  same result we would have obtained had we worked with a complex ${B}$,
  which shows that, in this case, the imaginary electric part is not harmful.  } so we get

\begin{equation}
\label{EuclideanAction3}
S_{E}(\aleph)
= 
\frac{g_{s}^{2}}{16\pi G_{N}^{(10)}}
\left\{\int_{\mathcal{M}} d^{10}x\sqrt{|{g}_{E}|}
\delta(\rho;\aleph)
+
2\int_{\partial\mathcal{M}}d^{9}x\sqrt{|h_{E}|}
e^{-2{\phi}}
\left[
2n_{\mu} 
\partial^{\mu}{\phi}
-(K-K_{0})
\right]
\right\}\, .
\end{equation}

Since the only non-compact coordinate is $\rho$, the boundary of $\mathcal{M}$
consists just of the asymptotic region $\rho=\rho_{\infty}$, where
$\rho_{\infty}$ is a regulator that must be taken to infinity eventually. In
the Euclidean NEOS configuration that we are considering, with the Euclidean
time compactified with the period $\beta=1/T_{H}$ the region $\rho=0$ is not a
boundary.

In the limit $\rho\rightarrow\infty$ 

\begin{equation}
n^{\mu}\partial_{\mu}{\phi}=\mathcal{O}(\rho^{-4})\, ,
\,\,\,\,\,
\mbox{and}
\,\,\,\,\,
K-K_{0}=\mathcal{O}(\rho^{-4})\, ,
\end{equation}

\noindent
so these terms decay too fast to contribute to the integral. Therefore, only
the bulk term gives a non-zero contribution and the Euclidean action is given
by 

\begin{equation}
S_{E}(\aleph)
=
\frac{g^{2}_{s}}{16\pi G_{N}^{(10)}}
V_{\mathrm{T}^{4}}V_{\mathrm{S}^{3}} 2\pi R_{y}\beta \Delta(\aleph)\, ,
\end{equation}

\noindent
where the 10-dimensional Newton constant is given in Eq.~(\ref{eq:GN}),
$V_{\mathrm{T}^{4}}=(2\pi l_{s})^{4}$ is the volume of the T$^{4}$,
$V_{\mathrm{S}^{3}}=2\pi^{2}$ is the volume of the S$^{3}$ of unit radius,
$R_{y}$ is the radius of the compact coordinate $y$ given in
Eq.~(\ref{eq:Ry}), $\beta=1/T_{H}$ is the period of the Euclidean time given
implicitly in Eq.~(\ref{eq:TH}) and $\Delta (\aleph)$ is the integral defined
in Eq.~(\ref{eq:Deltaa}). Substituting in the above expression, we find that

\begin{equation}
S_{E}(\aleph)
=
\frac{\pi R_{z}\sqrt{\tilde{Q}_{0} Q_{+}Q_{-}}}{2l_{s}^{4}}
\frac{\Delta(\aleph)}{\rho_{0}^{2}}\, .
\end{equation}

Note that this is a finite result for all the finite values of $\rho_{0}$. These
NEOS configurations are, therefore, instantons, even if they are not
solutions. We now have to undo the Wick rotation $\aleph=-i a$, so that
$\rho_{0}^{2}=\frac{1}{2}(Q_{+}^{2}-a^{2})^{1/2}$ and take the limit
$a\rightarrow Q_{+}$.

In this limit both $\Delta(a)$ and $\rho_{0}$ go to zero, but the limit
of $\Delta(a)/\rho_{0}^{2}$ turns out to be finite and takes the value

\begin{equation}
\lim_{a\rightarrow Q_{+}} \frac{\Delta(a)}{\rho_{0}^{2}}
=
g_{s}^{2}\log{\left(1+\frac{8\alpha'}{\tilde{Q}_{0}}\right)}
=
g_{s}^{2}\log{\left(1+\frac{8}{N_{S5}}\right)}
\approx
\frac{8g_{S}^{2}}{N_{S5}}\, ,
\end{equation}

\noindent
where we have used the assumption that $N_{S5}>> 8$.

This leads us to our final result for the Euclidean action associated to the
Lorentzian solution interpolating between two brane configurations we started
from

\begin{equation}
S_{E}=4\pi\sqrt{\frac{N_{F1}N_{W}}{N_{S5}}}\, ,
\end{equation}

\noindent
which leads to the transition probability 

\begin{equation}
|\mathcal{A}|^{2}\sim e^{-8\pi\sqrt{\frac{N_{F1}N_{W}}{N_{S5}}}}\, .
\end{equation}

\section{Discussion}
\label{sec-discussion}

First of all, we observe that, quite remarkably, this result coincides with
$e^{-\Delta S_{BH}}$, where $\Delta S_{BH}$ is the change of
Bekenstein-Hawking entropy in the process:

\begin{equation}
\Delta S_{BH}
=
2\pi\sqrt{N_{F1}N_{W}(N_{S5}+8)}-2\pi\sqrt{N_{F1}N_{W}N_{S5}}
\approx 
8\pi\sqrt{\frac{N_{F1}N_{W}}{N_{S5}}}\, ,
\end{equation}

\noindent
which is what one can expect on general grounds: for a single black hole, the
Euclidean action is proportional to the Bekenstein-Hawking entropy
\cite{Gibbons:1976ue,Kallosh:1992wa}\footnote{Again, we stress that, for
  extremal black holes, this calculation has to be made in the family of
  non-extremal black holes and then one has to take the extremal limit in the
  result to avoid the problems found in
  Refs.~\cite{Hawking:1994ii,Gibbons:1994ff}, as we have done here using a
  NEOS family.} and the Lorentzian solution we started from was interpreted as
connecting two different black-holes near-horizon geometries, so it is natural
that the Euclidean action yields the difference between the entropies of two
black holes.

Notice that the sign in the exponent is opposite to what one would expect from
statistical mechanics, where the decay rate would be estimated to be of the
order of $e^{+\Delta S}$. In our case, the entropy difference is positive, so
the process is favored thermodynamically. However, the decay process involves
topology change, which is highly suppressed.  This way we can interpret the
minus sign in $e^{-\Delta S}$ as the fact that topology constrains the decay
so effectively that it succeeds in suppressing the process even though it
involves an increase of entropy.

Given the composition of the string background corresponding to the black
hole (F1s, NS5s etc), the result can also be interpreted as the decay rate of
a gauge 5-brane which lives in a configuration of fundamental strings,
momentum waves and solitonic 5-branes.

When the numbers of each kind of component are comparable, this probability is
tiny and the gauge 5-brane, though unstable, is long lived. However, when the
number of S5-branes is much larger than the number of strings and waves,
$N_{S5}>N_{F1}N_{W}$, the gauge 5-branes decay quickly into S5-branes. One may
say that S5-branes are ``hungry'' for gauge 5-branes, and the larger their
number the faster they will eat them.

The result suggests that the non-Abelian 5-dimensional black hole
Eq.~(\ref{eq:bhsolution}) whose near-horizon limit gives the Lorentzian
``dumbbell'' solution we started from Eq.~(\ref{eq:nhsolution}) is
non-perturbatively unstable. Furthermore, and possible related to this fact,
it seems very difficult or, perhaps, it is impossible, to find non-extremal
black holes with the same charges and non-Abelian fields.  Clearly, more work
is necessary to clarify the situation.

Finally, we should comment on the relation between the result obtained here
and the method employed and Brill's work Ref.~\cite{Brill:1991rw} in which he
computed the Euclidean action of an instanton whose Lorentzian counterpart
connects several (at least three) asymptotic AdS$_{2}\times$ S$^{2}$
geometries and which has non-trivial topology.\footnote{These solutions are
  obtained from the Papapetrou-Majumdar family of solutions that describe
  Reissner-Nordstr\"om black holes in equilibrium \cite{kn:P,Majumdar:1947eu}
  with more than one center.  Removing the constant part of the harmonic
  function one finds an AdS$_{2}\times$ S$^{2}$ region in the near-horizon
  limits and another AdS$_{2}\times$ S$^{2}$ region at infinity whose charges
  are the sum of all those of the other regions. With just one center, one
  finds a single, global AdS$_{2}\times$ S$^{2}$ solution, with $N$ centers
  one find $N+1$ AdS$_{2}\times$ S$^{2}$ regions.} These solutions can be used
to study the non-perturbative splitting of a Reissner-Nordstr\"om black hole
into smaller black holes. In that case the Wick rotation offers no special
complications but the geometry of the instanton is very complicated and it is
not clear how to deal with the inner boundaries identified in
Refs.~\cite{Hawking:1994ii,Gibbons:1994ff}. The solution studied here is
topologically simpler (at least from the metric point of view), but we have
argued that a non-extremal off-shell (NEOS) deformation had to be used to go
to the Euclidean, compute the action and then go back to Lorentzian signature
and take the extremal limit, avoiding many of the pitfalls one finds along the
way.

We expect the NEOS technique developed here to be of further use in other
contexts.

\section*{Acknowledgments}

The authors would like to thank Patrick Meessen and Pedro F.~Ram\'{\i}rez for
many useful conversations.  This work has been supported in part by the
MINECO/FEDER, UE grant FPA2015-66793-P and by the Spanish Research Agency
(Agencia Estatal de Investigaci\'on) through the grant IFT Centro de
Excelencia Severo Ochoa SEV-2016-0597.  The work of P.A.C.~was supported by a
``la Caixa-Severo Ochoa'' International pre-doctoral grant. TO wishes to thank
M.M.~Fern\'andez for her permanent support.

\appendix


\end{document}